\documentclass[a4paper,twoside]{article}

\usepackage[utf8]{inputenc}
\usepackage{url}

\usepackage{epsfig}
\usepackage{subcaption}
\usepackage{calc}
\usepackage{amssymb}
\usepackage{amstext}
\usepackage{amsmath}
\usepackage{amsthm}
\usepackage{multicol}
\usepackage{multirow}
\usepackage{pslatex}
\usepackage{apalike}
\usepackage[bottom]{footmisc}
\usepackage{SCITEPRESS}     

\usepackage{siunitx}
\usepackage{comment}

\begin{document}

  \title{\huge Network Error Logging: HTTP Archive Analysis}

\author{\authorname{Kamil Jeřábek\sup{1}\orcidAuthor{0000-0002-5317-9222} and Libor Polčák\sup{1}\orcidAuthor{0000-0001-9177-3073}}
\affiliation{\sup{1}Brno University of Technnology, Faculty of Information
Technology, Božetěchova 2, 612 66 Brno, Czech Republic}
\email{\{ijerabek, polcak\}@fit.vut.cz}
}


  \abstract{
    Network Error Logging
    helps web server operators detect operational problems in real-time
    to provide fast and reliable services. HTTP Archive provides detail
    information of historical data on HTTP requests. This paper leverages the
    data and provides
    a long-term analysis of Network Error Logging deployment.
    The deployment
    raised from 0 to 11.73\,\% (almost 2,250,000 unique domains) since
    2019. Current deployment is dominated by Cloudflare. Although we observed
    different policies, the default settings prevail. Third party collectors
    emerge raising the diversity needed to gather sound data. Even so, many
    service deploy self-hosted services. Moreover, we identify potentially
    malicious adversaries deploy collectors on randomly-generated domains and
    shortened URLs.
}

\onecolumn \maketitle \normalsize \setcounter{footnote}{0} \vfill

\section{\uppercase{Introduction}}
Web server operators need to monitor their servers for availability, which is a
crucial success factor~\cite{franke2012}. Network Error Logging (NEL) is a
recent World Wide Web Consortium (W3C) proposed standard~\cite{w3c_nel} that
allows HTTP servers to employ browsers\footnote{At the time of the writing of
this paper, Chromium-based browsers like Google Chrome, Chrome for Android,
Microsoft Edge, Opera, and Opera GF support NEL. Brave is the only
Chromium-based browser that we discovered does not support NEL.} to report
failures in the reachability of web servers, including failures during domain
name resolution~\cite{paper_nel}.

Current literature provides only a very shallow data on NEL
deployment~\cite{nel_cdn} even though related work identified security and
privacy issues in the current NEL implementation~\cite{nelsecrypt23}. This paper
leverages HTTP archive and analyzes NEL deployment to shed more light on NEL
deployment and its trends. NEL deployment is rising, and its share is over
10\,\%.

This paper is the first to analyze the long-term deployment of NEL. Our analysis
shows that:

\begin{itemize}

  \item Current deployment is dominated by Cloudflare. Consequently, the NEL
  adoption share strongly depends on future decision of Cloudflare. This paper
  highlights Shopify, another NEL large provider, that stopped signalling NEL in
  2021.

  \item Even though the original NEL paper~\cite{paper_nel} suggests that Google
  was migrating to NEL, HTTP Archive data does not show any traces of Google
  deploying NEL.

  \item We provide data on the NEL policies deployed in the wild and study the
  NEL ecosystem and the trends inside.

  \item We identify NEL deployment on domains with strange, possibly randomly
  generated, names that employ collectors hidden behind a shortened URL.

\end{itemize}

This paper is organized as follows. Section~\ref{sec:nelbackground} explains how
NEL works. Section~\ref{sec:related} overviews related
work.
Section~\ref{sec:results} describes the methodology and results of the analyses
of the HTTP Archive of the NEL deployment.
Section~\ref{sec:discussion} discusses the results of this
  paper.
Section~\ref{sec:conclusion} concludes this paper.

\section{\uppercase{NEL Background}}
\label{sec:nelbackground}
\noindent NEL was introduced~ by researchers mostly affiliated with
Google~\cite{paper_nel}. The World Wide Web
Consortium (W3C) is in the process of standardizing
NEL~\cite{w3c_nel}.
NEL introduces {\tt NEL} HTTP header sent by an HTTP server that contains the NEL
policy of the server. In addition,
{\tt Report-To} HTTP header determines web servers that collect
NEL reports for the visited domain.

The goal of NEL is to let web server operators instruct their
visitors to report failures in attempting to visit the service as well as
successful visits.
A web server operator can control the fraction of failures
({\tt failure\_fraction})
and success reports ({\tt success\_fraction}), and other parameters in a NEL policy. The policy is valid
for a limited time ({\tt max\_age} parameter of a policy). By default, the policy
  applies  to the visited domains, but a server can also instruct the clients
  to apply the policy to subdomains ({\tt include\_subdomains} parameter of a policy).

Figure~\ref{fig:nel_report_w3c} shows an example of a NEL report. The client
reports the age of the error (the browser often sends messages with a delay)
and its type. Additionally,
the report contains other information about the event so that an
operator can react to the message.

\begin{figure}[h]
    \centering
\begin{verbatim}
{
  "age": 0,
  "type": "network-error",
  "url": "https://www.example.com/",
  "body": {
    "sampling_fraction": 0.5,
    "referrer": "http://example.com/",
    "server_ip": "2001:DB8:0:0:0:0:0:42",
    "protocol": "h2",
    "method": "GET",
    "request_headers": {},
    "response_headers": {},
    "status_code": 200,
    "elapsed_time": 823,
    "phase": "application",
    "type": "http.protocol.error"
  }
}
\end{verbatim}
    \caption{An example of a NEL report~\cite{w3c_nel}.}
    \label{fig:nel_report_w3c}
\end{figure}

As browsers store only the
last policy retrieved for each domain, the HTTP server should install the same
policy on all pages on each domain.

\section{\uppercase{Related Work}}
\label{sec:related}
Goenka et al.~\cite{nel_cdn} focused on using dynamic NEL configuration to show
how NEL can be used for active measurements critical to CDN operations such as
alternate PoP measurements. Besides the main topic, Goenka et al. also briefly analyzed
the HTTP Archive. This paper differs as (1) they analyzed only three samples in
2020 whereas this paper covers data between 2018 and 2023, (2) this paper
provides deep analysis of the deployed policies not available in their paper,
(3) they did not describe the deployment strategies and (4) did not foucus on
the collector diversity; although they report a rapid increase of reporting
URLs, they do not investigate how many domains operate these URLs.

Consequently, this paper provides much deeper analysis from the longitudinal
point of view as well as qualitative view.

Jeřábek and Polčák~\cite{nelsecrypt23} studied NEL from the data
protection perspective raising serious questions about the legality of NEL
deployment in the European Union. Additionally, they explore security issues
concerning NEL. The data on the NEL deployment in this paper sheds more light
on the seriousness of the issues.
To our best knowledge, all other papers concerning NEL
utilize NEL as a tool. This paper is the first that
systematically monitors the NEL deployment.

\section{\uppercase{HTTP Archive Analysis}}
\label{sec:results}
\noindent This section analysis NEL deployment to learn whether NEL
is actively deployed.

\subsection{Methodology}

HTTP Archive periodically crawls millions of URLs
at least once a month
and stores information about the fetched resources, including HTTP
headers~\cite{httparchive_methodology}.
The crawled data is publicly accessible for analysis via Google Cloud Big Query.
HTTP Archive has been crawling the web since November 2011. The data
from each crawl is organized in a separate table with a date label.
Besides the
original HAR logs\footnote{\url{https://docs.fileformat.com/web/har/}},
HTTP Archive provides summary tables with preprocessed records. A record is a preprocessed
HTTP request-response pair.
Each page visit during a crawl may include several HTTP requests and responses for additional resources
such as JavaScript, CSS, images, or other resources needed for the page load.
Hence, a single page visit is stored as multiple records.

We analyzed the summary tables since they contain all the necessary information.
We process all records of each page visit. However,
we only analyze the first record of each unique domain to remove duplicates
since we expect that operators deploy NEL policies consistently.

Moreover, HTTP Archive distinguishes mobile and desktop crawls. Some URLs appear
in both crawls. We investigated the overlapping
domains and checked that both crawls contain the same NEL policies. Hence, we
merged the mobile and desktop records and further analyzed overlapping domains
once.

We
base our analysis on data from HTTP Archive because it (1)~is publicly available, (2)~includes HTTP
headers, (3)~contains historical records, and (4)~contains a vast amount of visited websites. To our
best knowledge, we are not aware of any better data source providing a comparable amount of data on NEL
deployment.

We analyze the data from the 1st February desktop and mobile crawl in each of the
last six years (February 2018 to February 2023). We selected the day
because a crawl labeled with that day occurs in every examined year for both
mobile and desktop.
We extracted the data using SQL-like Big Query jobs\footnote{https://github.com/kjerabek/nel-http-archive-analysis} and further processed in Python.

\subsection{Results}
In 2018, no service deployed NEL as the first implementation of NEL appeared in
Chromium at the end of 2018. The adoption raised in the
following years. However, the share dropped a little in 2022.
Section~\ref{sec:discussion} discusses a possible explanation for the drop.
Table \ref{tab:nel-portions} contains the number of all domains
included in the HTTP Archive, the number and the share of domains that deployed NEL and responded with NEL policy.

\begin{table}[h]
    \centering
    \caption{Unique domains queried within each crawl and those responding with valid NEL headers.}
    \label{tab:nel-portions}
    \begin{tabular}{|c|r|r|r|}
        \hline
        \multicolumn{1}{|c|}{Date} & \multicolumn{1}{|c|}{Domains} & \multicolumn{1}{|c|}{NEL} & \multicolumn{1}{|c|}{NEL [\%]} \\
        \hline
        Feb 2018 & \num{1022970} & \num{0} & \num{0} \\
        Feb 2019 & \num{5707189} & \num{355} &  \num{0,01} \\
        Feb 2020 & \num{6636205} & \num{109483} & \num{1,65} \\
        Feb 2021 & \num{10147089} & \num{1004279} & \num{9,90} \\
        Feb 2022 & \num{10363447} & \num{960033} & \num{9,26} \\
        Feb 2023 & \num{19159613} & \num{2247233} & \num{11,73} \\
        \hline
    \end{tabular}
\end{table}

The analysis below examines only the valid NEL responses. We only consider the domains that deployed NEL. Other domains are not analyzed further.
We do not repeat the exact date of the crawls below and refer to each sample by its year only.

Our analysis revealed several known companies and organizations
using NEL.
\url{Booking.com} first deploys NEL in the 2020 records, and
Cloudflare first injected NEL headers in the 2021 records. Wikimedia
monitors its services like \url{Wikipedia.org} and first appears in the 2021
records. Reddit monitors
\url{reddit.com} and other language mutations in the 2022 and later records. Microsoft monitors
Office365 websites like \url{sharepoint.com} in the 2022 and 2023 records. Other companies
experiment with NEL. For example, a bank site \url{Raiffeisen.ch} signals NEL
on domain \url{boerse.Raiffeisen.ch}, \url{dell.com} signals NEL on
\url{dl.dell.com} and \url{downloads.dell.com}, {fastly-insights.com} deploys
NEL on some subdomains only. Facebook uses NEL on 
\url{www.beta.facebook.com} and reports to \url{www.beta.facebook.com}.

\subsubsection{NEL Collectors}
\label{sec:nel_collectors}

An operator of a web service may deploy their own NEL collector, or the operator may employ
a \emph{NEL collector as a service} (collector hosting)~\cite{paper_nel}. The
collector-hosting provider would typically process reports for many domains. Our
analyses revealed both collector hosting and private (self-hosted) collectors.

The original paper~\cite{paper_nel} deems the collector diversity crucial. It
favors the services that send NEL reports to a completely different domain to
collect NEL reports in more failure scenarios. For example,
\url{developer.valvesoftware.net} collects NEL at \url{reports.valve.net},
{www.yandex.com}, and various other Yandex domains report to \url{dr.yandex.net}
and \url{dr2.yandex.net}.
In contrast, others signal to the same domain name, like \url{fel.cvut.cz} and \url{www.staremesto.sk}.

In addition, we analyzed collector URLs and domain names and identified two strategies of NEL collector naming to
distinguish the origin. (1) NEL collectors have
different subdomain for each origin. An example of such a collector is
\url{report-uri.com}: domain \url{www.expobeds.com} reports to
\url{expobeds.report-uri.com}, whereas \url{mattferderer.com} reports to
\url{mattferderer.report-uri.com}. (2)~All reporting domains share the same NEL
collector domain name, and the
URL path or parameters encode an identifier to distinguish the reporting
domain. An example of such a service is \url{cloudflare.com}, which uses
136-characters-long Base64 encoded string identifiers
in URL parameters.

To provide fair statistics
that are not skewed by the two collector naming strategies,
we denote a \emph{NEL collector provider} as a set of NEL collectors with the
same second-level
domain (SLD). At the beginning of
2019, only 11 (mostly self-hosted) collector providers appeared; the number raised to 147
collectors in 2023,
see Table~\ref{tab:report-to}.

\begin{table}[h]
    \centering
    \caption{The count of NEL collector providers, the top four
    NEL collector providers for each analyzed year, and their share over the analyzed period.}
    \label{tab:report-to}
    \begin{tabular}{|c|c|l|r|}
        \hline
        \multicolumn{1}{|c|}{Year} & Count & \multicolumn{1}{|c|}{Top 4
        providers} &
        \multicolumn{1}{|r|}{Share [\%]}\\ \hline \hline
        \multirow{4}{*}{2019} & \multirow{4}{*}{11} & report-uri.com & \num{87.32} \\
        & & 3gl.net & \num{5.63} \\
        & & uriports.com & \num{2.54} \\
        & & seloc.club & \num{1.41} \\ \hline \hline

        \multirow{4}{*}{2020} & \multirow{4}{*}{36} & shopifycloud.com & \num{98.84}\\
        & & report-uri.com & \num{0.56} \\
        & & powerboutique.net & \num{0.23} \\
        & & fastly-insights.com & \num{0.07} \\ \hline \hline
        
        \multirow{4}{*}{2021} & \multirow{4}{*}{84} & cloudflare.com & \num{77.44} \\
        & & shopifycloud.com & \num{22.26} \\
        & & wikimedia.org & \num{0.12} \\
        & & report-uri.com & \num{0.09} \\ \hline \hline
        
        \multirow{4}{*}{2022} & \multirow{4}{*}{115} & cloudflare.com & \num{97.72} \\
        & & cafe24.com & \num{1.79} \\
        & & wikimedia.org & \num{0.13} \\
        & & report-uri.com & \num{0.10} \\ \hline \hline
        
        \multirow{4}{*}{2023} & \multirow{4}{*}{147} & cloudflare.com & \num{97.91} \\ 
        & & cafe24.com & \num{0.94} \\
        & & freshedge.net & \num{0.34} \\ 
        & & office.net & \num{0.33} \\ \hline
        
    \end{tabular}
\end{table}

In 2019, \url{report-uri.com}, a collector-hosting service, prevailed. In later years, big CDN providers
Shopify and Cloudflare appeared and deployed NEL on the websites they hosted.
Both utilize their own NEL collector. 
Cloudflare's collector has appeared in the reporting destinations of hundreds of
thousands of domains in the 2021 crawl and even increased to almost two million in 2023.

Some collector providers collect NEL reports from a single domain, whereas others are
shared between multiple domains, see Table~\ref{tab:report-to-domains}. The
number of providers shared between many domains is rising. For example,
\url{nelreports.net} first appeared in 2020, serving 26 domains; it raised to 57
in 2021, 73 in 2022, and 1177 in 2023. Nevertheless, the
number of providers employed by a small number of domains also rises
significantly.

\begin{table}[h]
    \centering
    \caption{The number of NEL collector providers that are employed by the given number of
  domains.}
    \label{tab:report-to-domains}
            \begin{tabular}{|r|r|r|r|r|r|}
        \hline
        \multicolumn{1}{|c|}{Domains} & \multicolumn{1}{|c|}{2019} & \multicolumn{1}{|c|}{2020} & \multicolumn{1}{|c|}{2021} & \multicolumn{1}{|c|}{2022} & \multicolumn{1}{|c|}{2023}\\
        \hline
1            & 3 & 16 & 48 & 52 & 74 \\
2            & 4 &  4 &  8 & 18 & 19 \\
3--10        & 2 &  4 & 11 & 20 & 26 \\
11--100      & 1 &  9 & 11 & 15 & 11 \\
101--1K      & 1 &  2 &  3 &  7 &  9 \\
More         & 0 &  1 &  3 &  3 &  8 \\

        \hline
    \end{tabular}
\end{table}

\subsubsection{NEL Policy Settings}

We analyzed the NEL policies in HTTP Archive dataset and studied \texttt{failure\_fraction}, \texttt{success\_fraction}, \texttt{include\_subdomains}, and
\texttt{max\_age} that appear in the wild.

In the beginning, several different values appeared in
\texttt{failure\_fraction} policies, which reflects
the early adoption and possibly experimentation with the newly introduced technology. In 2020, the values
were set to report mostly a small share of errors (10\,\% in 99\,\%
of cases). However, since 2021 almost all operators have been interested in all errors, which is
the default.
Only one domain in 2020 was not interested in failures at all. The desire to
collect failures is understandable, as failure detection is the main motivation
behind NEL. Even so, 21,606 domains collect only
10\,\% of the failures, and 5,429 domains collect 5\,\% of the failures or less
(in 2023). Table~\ref{tab:failure-fraction-occurence} shows the settings observed each year.

\begin{table}[h]
    \centering
    \caption{Observed failure fraction values. The symbol $\updownarrow$ denotes
    an open interval of values between range values on the row above and below.}
    \label{tab:failure-fraction-occurence}
    \begin{tabular}{|c|r|r|r|r|r|}
        \hline
        \multicolumn{1}{|c|}{Value} & \multicolumn{1}{|c|}{2019} & \multicolumn{1}{|c|}{2020} & \multicolumn{1}{|c|}{2021} & \multicolumn{1}{|c|}{2022} & \multicolumn{1}{|c|}{2023}\\
        \hline
0 & 0 & 1 & 0 & 0 & 0 \\
$\updownarrow$ & 224 & 297 & 363 & 458 & 3761 \\ 
0.01 & 1 & 108K & 78 & 107 & 137 \\
$\updownarrow$ & 0 & 0 & 0 & 1 & 5 \\
0.05 & 0 & 4 & 1269 & 1328 & 1526 \\
$\updownarrow$ & 0 & 0 & 0 & 0 & 0 \\ 
0.1 & 3 & 266 & 167 & 17K & 22K \\
$\updownarrow$ & 0 & 2 & 3 & 19 & 78 \\
0.25 & 0 & 0 & 5 & 6 & 8 \\
$\updownarrow$ & 0 & 0 & 0 & 0 & 2 \\
0.5 & 2 & 3 & 3 & 9 & 72 \\
$\updownarrow$ & 1 & 0 & 0 & 9 & 6 \\
1 & 124 & 673 & 1.0M & 941K & 2.2M \\
        \hline
    \end{tabular}
\end{table}

Regarding \texttt{success\_fraction}, we typically observed ranges from
\num{0.0} to \num{0.01}, as shown
in Table~\ref{tab:success-fraction-occurence}. The default value of \num{0.0} dominates in all years except for
2020 where the servers injected the policy of $0.001$ in most cases.
Although most servers limit the number of success reports,
in the 2020 and later crawls, at least two domains are
interested in all success reports in each analyzed year.

\begin{table}[h]
    \centering
    \caption{Observed success fraction values. The symbol $\updownarrow$ denotes
    an open interval of values between range values on the row above and below.}
    \label{tab:success-fraction-occurence}
    \begin{tabular}{|c|r|r|r|r|r|}
        \hline
        \multicolumn{1}{|c|}{Value} & \multicolumn{1}{|c|}{2019} & \multicolumn{1}{|c|}{2020} & \multicolumn{1}{|c|}{2021} & \multicolumn{1}{|c|}{2022} & \multicolumn{1}{|c|}{2023}\\
        \hline
0 & 350 & 1156 & 780K & 907K & 1.6M \\
$\updownarrow$ & 1 & 108K & 224K & 805 & 9267 \\
0.01 & 2 & 1 & 19 & 51K & 620K \\
$\updownarrow$ & 0 & 0 & 0 & 0 & 3 \\
0.05 & 0 & 14 & 7 & 21 & 41 \\
$\updownarrow$ & 0 & 0 & 0 & 0 & 0 \\
0.1 & 0 & 0 & 2 & 3 & 8 \\
$\updownarrow$ & 0 & 0 & 0 & 9 & 509 \\
0.25 & 2 & 2 & 3 & 0 & 0 \\
$\updownarrow$ & 0 & 0 & 0 & 0 & 0 \\
0.5 & 0 & 0 & 0 & 0 & 4 \\
$\updownarrow$ & 0 & 0 & 0 & 0 & 0 \\
1 & 0 & 2 & 6 & 4 & 19 \\
        \hline
    \end{tabular}
\end{table}

As NEL policies are not applied to subdomains by default, operators need to
intentionally change the policy to include subdomains. The non-default setting
prevailed only in the first year of the NEL deployment. Later, the policies apply to subdomains in
about or less than $1\%$ of cases. Table~\ref{tab:subdomains-true} shows the precise distribution in all years.

\begin{table}[h]
    \centering
    \caption{The share of NEL-signalling servers with {\tt include\_subdomains} set to true.}
    \label{tab:subdomains-true}
    \begin{tabular}{|c|r|}
        \hline
        \multicolumn{1}{|c|}{Year} & \multicolumn{1}{|c|}{true [\%]} \\ \hline
        2019 & 98.03 \\
        2020 & 1.02 \\
        2021 & 0.14 \\
        2022 & 0.17 \\
        2023 & 0.63 \\ \hline
    \end{tabular}
\end{table}

The expiration policy, set by \texttt{max\_age},
ranges from seconds to years since the maximal value is not limited. Nevertheless, the most common
values occurring are hours, 1 day, 7 days, and 30 days. Hundreds of domains inject policies
valid for one year. The maximal value observed was two years, but it appeared only
in a few cases. Some servers removing NEL policies with {\tt max\_age=0} appeared each year.
Table~\ref{tab:max-age-occurence} shows the distribution of \texttt{max\_age} values in each year.

\begin{table}[h]
    \centering
    \caption{Observed {\tt max\_age} values. The symbol $\updownarrow$ denotes an open
    interval of values between range values on the row above and below;
    \emph{ds}
    stands for days.}
    \label{tab:max-age-occurence}
    \begin{tabular}{|c|r|r|r|r|r|}
        \hline
        \multicolumn{1}{|c|}{Value} & \multicolumn{1}{|c|}{2019} & \multicolumn{1}{|c|}{2020} & \multicolumn{1}{|c|}{2021} & \multicolumn{1}{|c|}{2022} & \multicolumn{1}{|c|}{2023}\\
        \hline
0 & 1 & 5 & 9 & 6 & 7 \\
$\updownarrow$ & 3 & 42 & 207 & 318 & 3645 \\
1 hour & 228 & 302 & 466 & 1177 & 840 \\
$\updownarrow$ & 0 & 333 & 0 & 78 & 7485 \\
1 day & 5 & 50 & 1490 & 1562 & 2866 \\
$\updownarrow$ & 1 & 0 & 0 & 7 & 54 \\
7 ds & 3 & 62 & 778K & 939K & 2.2M \\
$\updownarrow$ & 2 & 68 & 76 & 94 & 87 \\
30 ds & 29 & 108K & 224K & 18K & 30K \\
$\updownarrow$ & 4 & 10 & 63 & 21 & 32 \\
365 ds & 79 & 234 & 486 & 489 & 881 \\
$\updownarrow$ & 0 & 2 & 2 & 8 & 9 \\
730 ds & 0 & 1 & 1 & 1 & 2 \\
        \hline
    \end{tabular}
\end{table}

\section{\uppercase{Discussion}}
\label{sec:discussion}
\noindent This section discusses the results and limitations of this paper.

\subsection{Observations From the HTTP Archive Analysis}

Since 2020 the
default policy values have prevailed.
However, non-default values also appear in the data set. We observed values for each variable across its
validity range.
The NEL deployment of Cloudflare causes the uniformity of the observed values.
We observed only two policies that send data to the NEL collector of Cloudflare that
differ in \texttt{success\_fraction} only.

We suspect that Shopify stopped signaling NEL headers during 2021 and we observed the drop in
NEL share in 2022. In 2023, over 97\,\% of NEL deployment is controlled by
a single entity --- Cloudflare. Hence, the observed 11.73\,\% share of NEL
deployment strongly depends on Cloudflare adoption. Should Cloudflare remove NEL,
NEL deployment share would dwindle.

We did not find any traces of Google's NEL deployment. Even when we manually visited
Google domains, we did not see NEL headers. The original NEL
paper~\cite{paper_nel} mentions two implementations. One is called \emph{Domain
Reliability} and was hard-coded in Chrome since 2014 (version 38). The other
is based on HTTP headers, as studied in this paper. The original NEL
paper~\cite{paper_nel} claims that Google was migrating from Domain Reliability
to NEL. However, we have not observed Google sending NEL headers; it looks like
Google never deployed NEL based on HTTP headers. Another explanation is that
Google signals NEL headers only to a fraction of visitors, and neither our manual
attempts nor the crawls of HTTP Archive triggered the injection.

In the last two years' records, we have encountered a few cases of randomly generated
domains, like \url{gkjw.org} and \url{embeumkm.com}. These domains were
injecting NEL policies with the reporting destination set to shortened URLs hosted by
\url{bit.ly}. The policy setting used default values, except it included subdomains
and lasted one year. Such an approach
demonstrates how malicious actors can misuse NEL to monitor their deployed services
while hiding their NEL collectors, as it is impossible to retroactively
learn the true identity of these collectors from HTTP Archive.

\subsection{Limitations of this study}

During the HTTP Archive analyses, we focused on the first response of each
domain. We did not analyze if the NEL header setting is consistent within a
domain. As HTTP Archive focuses on main pages and pages linked from the main
page~\cite{httparchive_methodology}, only a minority of resources
are crawled on most domains. Hence, a sound analysis of the consistency of the
NEL policies is not possible based on HTTP Archive. For these reasons and
because HTTP Archive does not crawl data from various locations, we cannot
confirm if NEL leaks data to third parties in practice.

The original NEL paper~\cite{paper_nel} recommends
hosting collectors in multiple physical and network (logical) locations.
However, data concerning physical and network locations are not available in the
HTTP Archive.
Consequently,
Section~\ref{sec:results} does not study some characteristics of NEL deployment,
like collector diversity.
Further work can amend Section~\ref{sec:results} with a study of additional
aspects of NEL deployment.

A limitation of the study of NEL collectors (subsection~\ref{sec:nel_collectors}) is that we classify NEL collectors
according to the second-level domain rather than the effective TLD+1. Consequently,
we aggregate all collectors in \url{co.uk} and \url{co.th} domains (both domains
have two collectors). Additionally, we aggregate tenants of some cloud providers
like Amazon AWS (\url{amazonaws.com}). We accepted the limitation because
(1)~the number of collectors in the TLDs without the possibility to register
second level is tiny so the results are not significantly affected and (2)~it is
not clear where to draw the line (should the Amazon AWS tenants be counted
separately or together; in the former case, what other services should be
treated similarly?).

\section{\uppercase{Conclusion}}
\label{sec:conclusion}
\noindent Error monitoring is a crucial activity of web server operators.
Even though Chromium-based browsers have supported NEL for several years, NEL
deployment was not systematically studied.
Our paper studies the current NEL deployment and revealed that NEL is dominated by
Cloudflare. Its customers employ one of two policies. Nevertheless, we observe
other important players that deployed NEL or experiment with the deployment.

\section*{\uppercase{Acknowledgments}}
This work was supported by the Brno University of
Technology grant Smart information technology for a resilient society
(FIT-S-23-8209).

\bibliographystyle{apalike}
{\small
\bibliography{main}}

\end{document}